\documentclass[fleqn,twoside]{article}
\usepackage{espcrc2,epsf}

\title{QCD, Symmetry Breaking and the Random Lattice}
 
\author{Saul D. Cohen
\address{Department of Physics, Columbia University, New York, NY, 10027, USA}
}

\begin{document}

\begin{abstract}

According to the Nielsen-Ninomiya No-Go theorem, the doubling of
fermions on the lattice cannot be suppressed in a chiral theory.
Whereas Wilson and staggered fermions suppress doublers with explicit
breaking of chiral symmetry, the random lattice does so by spontaneous
chiral symmetry breaking even in the free theory.  I present results
for meson masses, the chiral condensate and fermionic eigenvalues
from simulations of quenched QCD on random lattices in four dimensions,
focusing on chiral symmetry breaking.

\end{abstract}

\maketitle

\section{Introduction}

Due to the nature of the QCD fermionic action on a regular hypercubic
lattice, the momentum of a fermion at the edge of the Brillouin zone is
effectively zero. Since the naive action has no distinction between the
expected modes near \(p = 0\) and the doubler modes, the resulting theory
contains \(2^4\) fermions rather than just one. This problem may be resolved
by adding a Wilson term to the action, which increases for the doubler modes
as the lattice spacing \(a\) goes to zero. However, such a term explicitly
breaks the chiral symmetry of the theory, which may be restored using
computationally expensive Ginsparg-Wilson fermions.

Rather than eliminating doublers by attaching new terms to the action
near the edges of the Brillouin zone, suppose it were possible to eliminate
the edges of the Brillouin zone entirely. If the lattice were not periodic
at the scale \(a\), there would be no maximum value of momentum at \(\pi/a\).
An irregular lattice would have an infinitely large Brillouin zone, eliminating
the most obvious source of the doubler problem without introducing explicit
chiral symmetry breaking.

Toward this end, one might propose the calculation of QCD parameters on a
random lattice rather than a hypercubic one. In the following sections, we
describe the process by which such a lattice may be constructed, an
appropriate action for working on the random lattice, and the results of
small calculations on the random lattice. These calculations have
been done on four-dimensional random lattices with a SU(3) gauge field.

\section{Constructing the Random Lattice}

The construction of the four-dimensional random lattice (following \cite{CFL})
begins by randomly selecting the location of \(L^4\) points from a uniform
interval in each dimension. Since points that are located very close together
contribute huge amounts to the action (See Section \ref{sec:action}), and
thus become closely locked, we suppress the creation of such points by
excluding a small region around each from containing other points.

The random points selected specify a unique division of space into cells,
known as the Voronoi tessellation. Each cell contains all parts of space
that are closer to the associated point than to any other point on the
lattice. On the regular lattice, these cells are simply hypercubes of
side length \(a\), but on the random lattice they are complicated 4-d bodies.

The graph dual to the Voronoi tessellation is called the Delaunay
triangulation. Any two points whose Voronoi cells share a boundary are
connected by a link. Each link is dual to the 3-dimensional facets of a
cell in the same way that each point is dual to the cell itself. The links
also form higher-dimensional objects, such as triangles which are dual to
2-dimensional surfaces of the facets.

In a \(d\)-dimensional space, it may be shown that any mutually connected
set of \(d+1\) points (forming a \(d\)-simplex in the Delaunay graph) lie
on a circumsphere whose center is an intersection in the Voronoi tessellation.
This property allows us to construct the random lattice by the following
algorithm: First, find a five-point circumsphere that does not contain any
other points of the lattice. Add links among all these points. Then, using
an arbitrary subset of four points, find the point defining a new circumsphere
that does not contain any other points, and connect this point to the other
four. This process may be repeated until all points have been added to the
lattice.

For later use, we would also like to calculate the sizes of objects in the
lattice. Although it is easy to compute 1-d lengths and 2-d areas,
determining 3-d volumes and 4-d contents is a formidable challenge. For our
purposes, we have determined these values by Monte Carlo estimation.

\section{QCD on the Random Lattice}\label{sec:action}

Having constructed the random lattice, we would now like to specify a
method for performing calculations on it. We will use the action proposed
by Christ, Friedberg and Lee in \cite{CFL}:

\[
S_g = \frac{\beta}{6} \sum_{\triangle T}\frac{\tilde{A}_T}{2A_T}
\left(1-\frac{1}{3}\mathrm{ReTr}U_T\right)
\]

Rather than the usual product of links around a square plaquette, the
random lattice gluonic action involves products of links around the
fundamental triangles. Each triangle enters the action with a weight
proportional to the area of its dual and inversely proportional to its
own area. Thus, gauge fields will be allowed to have large deviations
from the identity on large triangles, but will be tightly constrained
on small triangles.

\begin{figure}[tb]
\epsfxsize=\hsize
\epsfbox{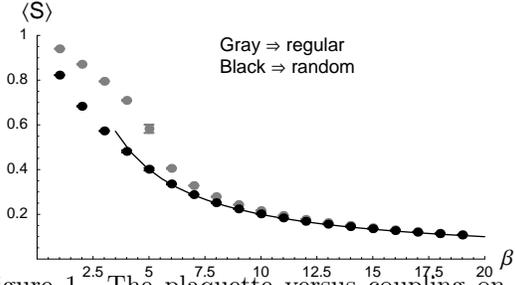}
\vspace{-0.5in}
\caption{The plaquette versus coupling on \(8^4\) regular and random
lattices.}
\vspace{-0.25in}
\end{figure}

In Figure 1, we plot the plaquette on random and regular lattices
versus \(\beta\). We see that the random lattice matches the regular
lattice in the weak-coupling limit, going as \(4/\beta\), but it has a
smoother approach to the strong coupling limit. There is no abrupt jump
between the two limits on the random lattice. (For earlier results, see
\cite{Ren}.)

The fermionic action enters similarly as a sum over links and points:

\[
S_f = \frac{1}{2}\sum_{\rightarrow L}
\bar{\psi}(x+l)\frac{\tilde{V}_L}{l_L}
(l_L^\mu \gamma_\mu)U_L\psi(x) \]\[
+ m_q \sum_{\cdot P} \tilde{\omega}_P\bar{\psi}(x)\psi(x)
\]

The mass term is weighted heavily for points associated with large cells,
and the covariant derivative is weighted heavily for short links and links
associated with large facets.

The Dirac operator, \(\mathcal{D}_{i,j}\), is derived from this action, but
the weights due to the geometry of the lattice are properly associated with
the final integral over spacetime rather than the operator, so they are
stripped out as in \cite{ERGW}. Where no link exists between points \(x_i\)
and \(x_j\), \(\tilde{V}_{i,j}\) should be considered zero.

\[
\mathcal{D}_{i,j} = 
\frac{\tilde{V}_{i,j}}
{2l_{i,j}\sqrt{\tilde{\omega}_i\tilde{\omega}_j}}
(l_{i,j}^\mu \gamma_\mu)U_{i,j}
+ m_q \delta_{i,j}
\]

The regular lattice may be expressed exactly as the limit in which the
points of the random lattice are forced to be in the usual hypercubic
arrangement. In this limit, the extra links created by triangulation have
zero weight in the Dirac operator. We may therefore create a continuum of
quasi-regular lattices interpolating between regular and random lattices by
allowing points to deviate from their regular lattice positions by \(a/2^n\).

\section{Observables on the Random Lattice}

By checking the expectation value of the Dirac Laplacian on plane waves,
we may see that the Brillouin zone of the random lattice is indeed infinite.
Rather than returning to zero at momentum \(\pi/a\), the value of the
Laplacian remains large.

\begin{figure}[tb]
\epsfxsize=\hsize
\epsfbox{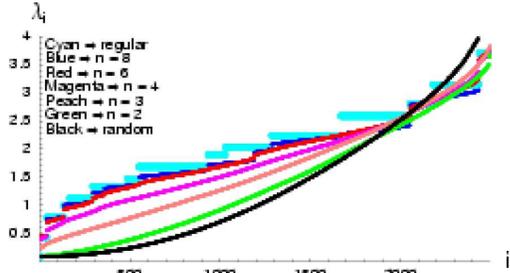}
\vspace{-0.5in}
\caption{Eigenspectra of the Dirac Laplacian for \(5^4\) lattices of varying
randomness.}
\vspace{-0.25in}
\end{figure}

But how can the elimination of doublers be reconciled with the Nielsen-
Ninomiya No-Go theorem, which indicates that chiral symmetry is
inconsistent with a non-doubled theory? In Figure 2, we examine the
eigenspectrum of the Dirac Laplacian on a range of quasi-regular lattices
(parametrized by \(n\)). In the region \(2<n<3\) (between the second-
darkest and third-darkest curves) we note that the character of the
spectrum near zero changes markedly. A large number of zero modes appear
on a sufficiently random lattice.

Examination of quark bilinear propagators shows a similar change of
character in this region. Sufficiently random lattices (with non-zero
coupling to the gauge fields) have a heavy vector bilinear, whereas more
regular lattices have a light vector. (See the open circles on Figure 3.)

As a function of coupling, we can also see the difference between the
two classes of lattices.  In Figure 3, the mass of the pseudoscalar
Goldstone (filled circles) is independent of the coupling, since the data
for all three values of the coupling lie on top of each other. This is in
sharp contrast to the regular lattice pseudoscalar Goldstone, which is
affected by deconfinement at weak coupling.

Together, these results indicate a spontaneous breaking of chiral
symmetry on the random lattice. The broken symmetry supplies a large
number of near-zero modes in the Dirac spectrum and creates a type
of Goldstone particle that is not associated with the symmetry breaking
of the QCD vacuum. Thus, the random lattice's pseudoscalar masses are
affected solely by the geometry of the lattice and are insensitive to
adjustments to the coupling constant.  Furthermore, the breaking of
chiral symmetry by the geometry of the random lattice produces
large additive fermion mass renormalizations, of \(\mathcal{O}(1/a)\),
as revealed by the heavy vector masses in Figure 3 (and suggested in
\cite{PW}).

\begin{figure}[tb]
\epsfxsize=\hsize
\epsfbox{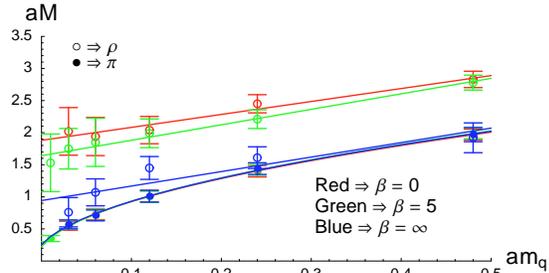}
\vspace{-0.5in}
\caption{Quark bilinear masses versus quark mass for various couplings
on a \(7^4\) random lattice.}
\vspace{-0.25in}
\end{figure}

\section{Conclusions}

Although the random lattice does appear to solve the doubling problem,
it is wholly unsuitable for computing realistic QCD properties. The
physics of the random lattice is dominated by pseudoscalar Goldstones
induced by a breaking of chiral symmetry associated with the geometry
of the lattice rather than the gluonic fields. Further, when the
coupling to the gauge fields is non-zero, all other physics is obscured
by the large additive fermion mass renormalization which is due to the
presence of the chiral condensate.  Of course, these problems must
disappear as both \(a\) and \(g\) are taken to zero, but at realizable
values of these parameters, the random lattice will remain far from
correspondance with ordinary QCD.

\end{document}